\documentclass[10pt,journal,twocolumn]{IEEEtran}

\usepackage{amsfonts}
\usepackage{cite}
\usepackage{graphicx}
\usepackage{psfrag}
\usepackage{amsmath}
\usepackage{times}
\usepackage[dvips]{epsfig}
\usepackage{amssymb}
\usepackage{subfigure}
\usepackage[center,small]{caption2}
\usepackage[below]{placeins}

\usepackage{url}
\usepackage{stfloats}

\usepackage{array}
\usepackage[linesnumbered]{algorithm2e}

\usepackage[table]{xcolor}
\usepackage{multirow}

\begin{document}

\title{Pairwise Check Decoding for LDPC Coded Two-Way Relay Block Fading Channels}

\author{Jianquan Liu, Meixia Tao, \emph{Senior Member, IEEE}, and Youyun Xu, \emph{Senior Member, IEEE}
\thanks{This work was partly presented in IEEE ICC 2010~\cite{ICC10:We} and IEEE ICC 2011~\cite{ICC11:We}.}
\thanks{The authors are with the Department of Electronic Engineering, Shanghai Jiao Tong University, Shanghai, P. R. China, 200240. Youyun Xu is also with the Institute of Communications Engineering, PLA University of Science \& Technology, Nanjing,
P. R. China, 210007. Email: jianquanliu@sjtu.edu.cn, jianquanliu@gmail.com, mxtao@sjtu.edu.cn, yyxu@vip.sina.com.}
\thanks{This work is supported in part by the National 973 project under grants 2012CB316100 and 2009CB320402, the National Natural Science Foundation of China under grants 60902019 and 60972050, and the Innovation Program of Shanghai Municipal Education Commission under grant 11ZZ19.}
}

\maketitle

\IEEEpeerreviewmaketitle

\begin{abstract}

Partial decoding has the potential to achieve a larger
capacity region than full decoding in two-way relay (TWR)
channels. Existing partial decoding realizations are however
designed for Gaussian channels and with a static physical layer
network coding (PLNC). In this paper, we propose a new solution for
joint network coding and channel decoding at the relay, called \emph{pairwise check
decoding} (PCD), for low-density parity-check (LDPC) coded TWR
system over block fading channels. The main idea is to form a check
relationship table (check-relation-tab) for the superimposed LDPC
coded packet pair in the multiple access (MA) phase in conjunction
with an adaptive PLNC mapping in the broadcast (BC) phase. Using
PCD, we then present a partial decoding method, two-stage
closest-neighbor clustering with PCD (TS-CNC-PCD), with the aim of
minimizing the worst pairwise error probability. Moreover, we propose the minimum correlation
optimization (MCO) for selecting the better
check-relation-tabs. Simulation results confirm that the proposed
TS-CNC-PCD offers a sizable gain over the conventional XOR with
belief propagation (BP) in fading channels.

\end{abstract}



\section{Introduction}

Two-way relaying, where two source nodes exchange information with
the help of a relay node, has recently gained a lot of research
interests~\cite{CISS05:Wu, VCT06:Larsson, VTC06:Popovski,
ICC06:Hausl, AMC06:Zhang, ASC06:Katti, JSAC07:Rankov, TSP10:Song, TVT10:Song, TSP12:Wang}. It is shown
able to overcome the half-duplex constraint and significantly
improve the system spectral efficiency in relay-based cooperative
networks. Upon receiving the bidirectional information flows, the
relay node combines them together and then broadcasts to the two
desired destinations. The operation at the relay resembles network
coding~\cite{IT00:Ahlswede}, a technique originally developed for
wireline networks. It is thus often referred to as physical layer
network coding (PLNC)~\cite{AMC06:Zhang} or analog network coding
(ANC)~\cite{ASC06:Katti}.

Among the various two-way relaying strategies, the two practical and
efficient ones are known as amplify-and-forward (AF) and
decode-and-forward (DF), similar to those in one-way relaying.
Different from one-way relaying, the DF strategy for two-way
relaying further includes full DF~\cite{IT08:Oechtering, IT08:Kim}
and partial DF~\cite{ITW08:Schnurr, CCC08:Gunduz, arxiv09:Kim}. This
is because the combining process operated at the relay is a
many-to-one mapping (e.g. $1 \oplus 1 = 0 \oplus 0 =0$). As a
result, it is not necessary for the relay to fully decode the
message pair before combining them together. Being a simple
realization of partial DF, the denoise-and-forward (DNF) strategy
proposed in \cite{ICC07:Popovski} demonstrates significant
performance gain over the full DF. It is also known that partial
decoding has the potential to achieve a much larger rate region than
full decoding, even though its capacity region is still unknown.
Consequently, it remains as a fundamental and challenging task to
realize the potential of partial decoding in two-way relay (TWR)
channels through practical coding and modulation techniques.

Several works have been reported on the implementations of partial
DF for channel-coded TWR systems, and they are also known as joint
network-coding and channel-coding (JNCD) design in the literature.
An intuitive method is to utilize the fact that the network-coded
(eg. XOR or modulo addition) codeword pair is also a valid codeword
given the same linear code (e.g. LDPC or Lattice codes) applied at
both source codes~\cite{IZS08:Nam, ITW07:Nazer, ACCCC07:Narayanan,
IT10:Wilson}. In this method, the relay first computes the
probability of $x_A \oplus x_B$ based on the received superimposed
signal $y_C$ during the multiple-access (MA) phase, where $x_A$ and
$x_B$ are the symbols after channel coding and modulation from
source $A$ and source $B$, respectively, and then apply
soft-decoding to decode the associated network-coded information
symbol pair $s_A \oplus s_B$. We refer to this method as partial DF
based on conventional XOR. However, this method discards useful
information related to the decoding of the whole packet $s_A \oplus
s_B$ during the mapping from signal $y_C$ to the probability of $x_A
\oplus x_B$. A more advanced partial decoding method is to exploit
the Euclidean distance profile of the superimposed symbol pair after
going through the noisy channel in the MA phase~\cite{JSAC09:Zhang}.
The relay first decodes the arithmetic-sum of the coded symbol pair
$c_A + c_B$ and then map it to $s_A \oplus s_B$ for broadcasting.
The authors in~\cite{JSAC09:Zhang} show that this partial DF method
based on arithmetic-sum provides higher decoding gain than the one
based on conventional XOR. Note that both aforementioned methods are
designed specifically for symmetric and Gaussian channels.

For TWR channel with fading, the conventional XOR does not always
work well due to the undesired phase and amplitude offset between
the two TWR channels in the MA phase. Authors
in~\cite{JSAC09:Koike-Akino} therefore proposed an adaptive network
coding with respect to the instantaneous channel fading, named as
closest-neighbor cluster (CNC) mapping. Compared to the conventional
XOR, the CNC mapping obtains a higher end-to-end throughput. To
further ensure reliable communication, the authors extended this
method for convolutional-coded system in~\cite{ICC09:Koike-Akino}
and discussed the code design based on trellis-coded modulation
(TCM)~\cite[Section 8.2]{BOOK01:Burr}. However, this TCM-based CNC
mapping requires to change the coding structure at the two source
nodes and switches two transmission protocols (CNC DNF and pseudo
AF) in order to exploit the system performance.

In this paper, we propose a new relay channel decoding solution,
called \emph{pairwise check decoding} (PCD), for LDPC coded TWR
fading channels. The main idea is to form a check relationship table
(check-relation-tab) for the coded symbol pair $(c_A, c_B)$ by
taking both the employed LDPC codes and the adaptive PLNC mapping
into accounts. The proposed PCD algorithm is universal for any
adaptive PLNC mapping and does not require the LDPC codes at the two sources being identical.
It offers a practical and efficient
approach to realizing the promising DNF TWR strategy with advanced channel
coding. With the aim of maximizing the minimum Euclidean distance (MED)
between any two codewords, we present a partial decoding method,
two-stage CNC with PCD (TS-CNC-PCD), for TWR fading channels.
The proposed TS-CNC-PCD is appropriate for any choice of constellation size.
Moreover, a kind of correlative rows optimization, named as the minimum
correlation optimization (MCO), is proposed for selecting
the better check-relation-tabs. Simulation results confirm that the
proposed TS-CNC-PCD has significant coding gains over the
conventional XOR with belief propagation (BP) decoding algorithm.

The rest of the paper is organized as follows. In Section II, we
present the channel coding model for TWR block fading channels.
Section III analyzes the lower bound of the outage probability. The
design criterion of adaptive codeword mapping is interpreted in Section IV.
In Section V, we propose the PCD algorithm in detail. Section VI
present a two-stage CNC mapping based on PCD decoding, named as TS-CNC-PCD, in terms of optimizing the MED. The
convergence behaviors and the coding gains of the proposed TS-CNC-PCD
method are simulated in Section VII. Finally, we conclude the
paper in Section VIII.

\section{Channel coding model for two-way relay fading channels}
\begin{figure*}[t]
\centering
\includegraphics[width=5.2in]{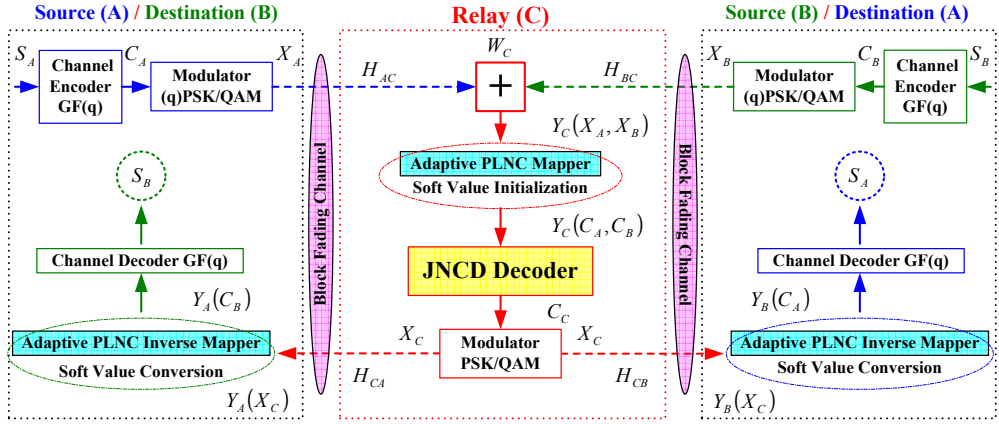}
\caption{Channel coding model for TWR block fading channels.}
 \label{fig:system_model}
\end{figure*}

We consider a TWR fading channel where two source nodes, denoted as
$A$ and $B$, exchange information with the help of a relay node,
denoted as $C$. We assume that all the nodes operate in the
half-duplex mode. The channel on each communication link is assumed
to be corrupted with block fading and additive white Gaussian noise
(AWGN). For simplicity, we also assume the channel gains are
reciprocal and unchanged during a whole packet transmission.

The proposed channel coding model is shown in
Fig.~\ref{fig:system_model}, where the communication takes place in
two phases. First, the information packet from each source, denoted
as $\bold{S}_i$, for $i\in \{A, B\} $, is encoded individually by a
traditional LDPC code with parity check matrix $\bold{\Gamma}_i$. Unlike
the existing work, we do not impose the constraint that $\bold{\Gamma}_A$
and $\bold{\Gamma}_B$ must be identical. Instead, we only require that
they have the same size and the same location of non-zero elements.
We further assume that the encoder is operated in $\bold{GF}(q)$,
where $q \in \{2^1,2^2,2^3,\ldots\}$. Note that $q$-ary ($q > 2$)
coding can improve the performance compared with binary coding. More
details can be found in~\cite{TC07:Declercq, TC08:Poulliat, IT09:Li}
and references therein. The encoded packet, $\bold{C}_i$, is
modulated by using $q$-ary modulation, denoted as ${\cal Q}_q$, such as $q$-PSK or $q$-QAM,
generating $\bold{X}_i$, and then transmitted simultaneously to the
relay node. The $n$-th symbol of each packet is denoted as
$\bold{C}_i(n)\in{\cal Z}_q$, ${\cal Z}_q = \{0,1,\ldots,q-1\}$, and
$\bold{X}_i(n) \in {\cal Q}_q$, respectively. The superimposed
packet received by the relay, denoted as $\bold{Y}_C$ is given by
\begin{equation}
\label{eq:model_1} \bold{Y}_C = H_{AC}\bold{X}_A + H_{BC}\bold{X}_B
+ \bold{W}_C,
\end{equation}
where $H_{ii'}$ denotes the complex-valued channel coefficient of
link from node $i$ to node $i'$, and $\bold{W}_{i'}$ denotes complex
AWGN with variance $\sigma_{i'}^2$ of node $i'$. Therein,
$i,i' \in \{A, B, C\}$.

We assume perfect symbol synchronization at the two sources and
perfect channel estimation at the relay. After receiving the
superimposed packet, the relay first computes the soft information (e.g. likelihood value) about the codeword after PLNC mapping, denoted as $\bold{C}_C = {\cal M}(\bold{C}_A, \bold{C}_B)$, then applies the proposed PCD to obtain the hard-decision (the details will be presented in Sections V and VI). Here, ${\cal M}$ denotes a kind of adaptive PLNC mapping, which is designed on the codeword pair, and $\bold{C}_C(n) \in {\cal Z}_{q'}$, where $q'$ is the cardinality of the PLNC mapped symbol $\bold{C}_C(n)$, which should be $q \leq q' \leq q^2$. Note that no full decoding of $\bold{C}_A$ and $\bold{C}_B$ is needed as
an intermediate step. No extra channel encoding at the relay is
needed either. Then, the relay broadcasts the modulated coded
symbols of $\bold{C}_C$, denoted as $\bold{X}_C$, and mapping rule
${\cal M}(\cdot)$ to two source nodes. The received signals at the nodes
$A$ and $B$ are respectively written as
\begin{equation}
\label{eq:model_2}
\begin{array}{l}
\bold{Y}_A = H_{CA}\bold{X}_C + \bold{W}_A;
\\ \bold{Y}_B = H_{CB}\bold{X}_C + \bold{W}_B.
\end{array}
\end{equation}

Each source node, say $B(A)$, computes the likelihood of the desired information
$\bold{C}_A(\bold{C}_B)$ from the received symbols
$\bold{Y}_B(\bold{Y}_A)$ by using the inverse PLNC mapping rule
with the help of its self-information $\bold{C}_B(\bold{C}_A)$.
Lastly, the traditional LDPC decoding algorithm, e.g. BP, is
applied, the output of which is the desired information packet
$\bold{S}_A(\bold{S}_B)$. Note that each source node should know the
check matrix of the other source.

\section{Analysis of outage probability}

In this section, we derive the system outage probability of TWR
fading channel, which serves as a good approximation of the
achievable frame error rate (FER) in the limit of infinite block
length~\cite{IT98:Biglieri}. Here, the system is said to be in
outage if the achievable sum-rate falls below a target. Since the
capacity region of two-way relaying with partial decoding is still
unknown~\cite{ITW08:Schnurr, CCC08:Gunduz, arxiv09:Kim}, we resort
to the capacity outer bound as follows~\cite[Theorem 2]{IT08:Kim},
based on which a lower bound of the outage probability can be
obtained.
\begin{equation}
\label{eq:outage_rate_region}
\begin{array}{l} \Lambda: \Big\{R_{AB}\leq
\min\big(\beta C_{AC}, (1-\beta)C_{CB}\big), \\ R_{BA}\leq
\min\big(\beta C_{BC}, (1-\beta)C_{CA}\big) \Big\},
\end{array}
\end{equation}
where $\beta$ is the time sharing parameter, $R_{ij}$ and $C_{ij}$
denote the instantaneous data rate and channel capacity of the link
from node $i$ to node $j$, for $i,j \in \{A,B,C\}$, respectively.

After simple manipulation of the constraints in
(\ref{eq:outage_rate_region})\footnote{The detailed derivation is similar to that described in~\cite{CC09:We}.}, we obtain the following linear
inequalities about $R_{AB}$ and $R_{BA}$
\begin{equation}
\label{eq:outage_linear_inequalities}
\begin{array}{l}
\frac{R_{AB}}{C_{AC}} + \frac{R_{AB}}{C_{CB}} \leq 1,
\\\frac{R_{AB}}{C_{AC}} + \frac{R_{BA}}{C_{CA}} \leq 1,
\\\frac{R_{AB}}{C_{CB}} + \frac{R_{BA}}{C_{BC}} \leq 1,
\\\frac{R_{BC}}{C_{BC}} + \frac{R_{BC}}{C_{CA}} \leq 1.
\end{array}
\end{equation}

Let us further assume that the TWR channels considered here are
reciprocal, i.e. $C_{ij} = C_{ji}$ for $i,j \in \{A,B,C\}$. From
(\ref{eq:outage_linear_inequalities}), we can easily obtain the
upper bound of the maximum sum-rate for the considered TWR channels
\begin{equation}
\label{eq:outage_sum_rate} S_u = \max\limits_{(R_{AB},R_{BA})\in
\Lambda} R_{AB} + R_{BA} =
\min(C_{AC},C_{BC}),
\end{equation}
which is also given in~\cite{ICC07:Popovski}. Therein, each of the
terms $C_{ij}$, $i,j \in \{A,B,C\}$, is the channel capacity of a
traditional point-to-point channel with input alphabet $x_{ij} \in
{\cal Q}_q$ and received signal $y_{ij} = \alpha_{ij} x_{ij} +
w_{ij}$, where $w_{ij} \sim {\cal N}(0,\sigma^2)$ and $\alpha_{ij}$
denotes a real- or complex-valued channel coefficient of the link
from node $i$ to $j$ with $\mathbb{E}\{|\alpha_{ij}|^2\}=1$.

With the further assumption of equiprobable channel inputs,
extending the well-known formula for the capacity of
continuous-valued Gaussian channels~\cite[Eqs. 3-5]{IT82:Ungerboeck}
to the case of block fading channels yields
\begin{equation}
\label{eq:outage_capacity}
\begin{array}{lll} C_{ij}(\alpha_{ij}) = \log_2(q)-
\frac{1}{q} \sum \limits_{m=0}^{q-1} \mathbb{E} \Bigg\{ \\ \log_2 \sum
\limits_{n=0}^{q-1} \exp
\bigg[-\frac{|y_{ij}-\alpha_{ij}x_{ij}^n|^2-|y_{ij}-\alpha_{ij}x_{ij}^m|^2}{2\sigma^2}
\bigg] \Bigg\}
\end{array}
\end{equation} in bit/channel use. Here, $\mathbb{E}$ represents expectation over $y_{ij}$ given $x_{ij} = x_{ij}^m$ and
$\alpha_{ij}$, with $x_{ij}^m$ being an element of the modulated
signal sets $\Big\{{\cal Q}_q:x_{ij}^0, x_{ij}^1, \dots,
x_{ij}^{q-1}\Big\}$.

In addition, we denote the target sum-rate of overall system as $S_r$. In our considered channel coded TWR model, given that the same LDPC code rate, denoted as $r$, and the same constellation are employed by the two source nodes, we have
\begin{equation}
S_r = r\log_2({q}).
\end{equation}
Then the outage probability
can be lower bounded as
\begin{equation}\label{eq:outage_lower_bound}
\begin{array}{lll}
 P_{out} \geq P(S_u < S_r) \\=P\Big(\min\Big\{C_{AC}(\alpha_{AC}),C_{BC}(\alpha_{BC})\Big\} < S_r\Big),
\end{array}
\end{equation}
which can be easily evaluated by Monte Carlo averaging over the
block fading coefficients and the AWGN.

\section{Design criterion of ${\cal M}(\bold{C}_A, \bold{C}_B)$}

Similar to the uncoded system in~\cite{JSAC09:Koike-Akino}, a
necessary condition for successful decoding at two
sources/destinations in coded TWR system is for the adaptive PLNC
mapping to satisfy the exclusive law:
\begin{equation}\label{eq:exclusive_law}
\begin{array}{lll}
{\cal M}(\bold{C}_A,\bold{C}_B) \neq {\cal
M}(\bold{C}_A',\bold{C}_B), \text{for all} \{\bold{C}_A \neq
\bold{C}_A',\bold{C}_B\}, \\
{\cal M}(\bold{C}_A,\bold{C}_B) \neq {\cal
M}(\bold{C}_A,\bold{C}_B'), \text{for all} \{\bold{C}_A,\bold{C}_B
\neq \bold{C}_B'\}.
\end{array}
\end{equation}

Given the above necessary condition, we next discuss the design
criterion of ${\cal M}(\bold{C}_A, \bold{C}_B)$ to optimize the system error
performance. We aim at minimizing the pairwise error probability
(PEP) between two distinct codewords $\bold{C}_C^l$ and
$\bold{C}_{C}^{l'}$ in the MA phase because the MA interference
dominates the whole system performance. Therein, $\bold{C}_C^l$ and
$\bold{C}_{C}^{l'}$ denote the corresponding codewords generated by the $l^{th}$ and $l'^{th}$ codeword pairs $(\bold{C}_A,\bold{C}_B)$ with certain selected PLNC mapping ${\cal M}(\cdot)$ respectively. Note that the PEP considered in this work is defined over codewords, whereas~\cite{JSAC09:Koike-Akino} was concerned with the PEP over uncoded symbols. For TWR fading channels, the PEP between
$\bold{C}_C^l$ and $\bold{C}_{C}^{l'}$ conditioned on the instantaneous channel gain pair
$\{H_{AC},H_{BC}\}$ is given by~\cite[p.
265]{BOOK65:Wozencraft}~\cite[Section 5.5]{BOOK01:Burr}~\cite[Eq.
7]{IT00:Knopp}~\cite[Eqs. 5-6]{JSAC09:Koike-Akino}
\begin{equation}
\label{eq:criterion_1}
\begin{array}{lll} P_e\big( \bold{C}_C^l \rightarrow
\bold{C}_C^{l'} \mid \{H_{AC},H_{BC}\} \big) \\= P_e\Big( {\cal
M}(\bold{C}_A^l,\bold{C}_B^l) \rightarrow {\cal
M}(\bold{C}_A^{l'},\bold{C}_B^{l'}) \mid \{H_{AC},H_{BC}\} \Big)
\\ = Q \bigg( \sqrt{\frac{ \zeta_{C} D^2_{ll'}
\mid \{H_{AC},H_{BC}\} }{2 \sigma^2}} \bigg),
\end{array}
\end{equation}
where $\zeta_{C}$ and $\sigma^2$ are the energy per coded symbol and
the variance of Gaussian noise at relay node C, respectively. $Q(\cdot)$ is the $Q$-function.
$D^2_{ll'}$ represents the squared Euclidean distance between the
codewords $\bold{C}_C^l$ and $\bold{C}_{C}^{l'}$ and is the function
of two channel coefficients $\{H_{AC},H_{BC}\}$ given by
\begin{equation}\label{eq:criterion_2}
D^2_{ll'} =  \sum \limits_{n=1}^{N}\Big|\hat{x}_{l}(n) -
\hat{x}_{l'}(n)\Big|^2.
\end{equation}
In (\ref{eq:criterion_2}), $N$ denotes the length of transmitted
codewords and
\begin{equation}\label{eq:criterion_3}
\begin{array}{lll}
\hat{x}_{l}(n) = H_{AC} x_A^{l}(n) + H_{BC} x_B^{l}(n),\\
\hat{x}_{l'}(n) = H_{AC} x_A^{l'}(n) + H_{BC} x_B^{l'}(n).
\end{array}
\end{equation}

From (\ref{eq:criterion_1}), it is clear that to minimize the
pairwise error probability, one should design a mapping rule
$\bold{C}_C = {\cal M}(\bold{C}_A, \bold{C}_B)$ that can maximize
the MED between any pair of codewords ($\bold{C}_C^{l},
\bold{C}_C^{l'}$). We denote it as
\begin{equation}\label{eq:criterion_4}
E^2 = \min \limits_{{\cal M}(\bold{C}_A^l,\bold{C}_B^l) \neq {\cal
M}(\bold{C}_A^{l'},\bold{C}_B^{l'})} D^2_{ll'}.
\end{equation}
Note that the optimal strategy of the mapping rule is adaptive with respect to the channel conditions.

\section{Pairwise check decoding (PCD)}\label{PCD}

After introducing the design criterion of the adaptive PLNC mapping $\cal{M}(\cdot)$ for coded TWR systems in the previous section, we present a general relay decoding framework, named as pairwise check decoding, for any given $\cal{M}(\cdot)$ in this section.

\subsection{Check relationship table (check-relation-tab) at relay}\label{PCD_table}

It is clear that if the conventional XOR mapping is applied, we can easily construct a virtual parity check matrix $\bold{\Gamma}_C$ for the codeword $\bold{C}_C$ based on $\bold{\Gamma}_A$ and $\bold{\Gamma}_B$ at the relay and then decode $\bold{C}_C$ using the traditional BP algorithm~\cite{IT62:Gallager,IT01:Richardson}. It is because the referred XOR mapping is linear. However, if an adaptive PLNC mapping is applied, which is much more likely to be non-linear, such virtual parity check matrix $\bold{\Gamma}_C$ in explicit form cannot be found. Instead, one has to resort to the constraint relationship regarding the codeword pair. In this subsection, we introduce a so-called check-relation-tab to describe such codeword
pair constraints.

At first, we set forth some notations. We assume that two LDPC codes
$\bold{\Gamma}_A$ and $\bold{\Gamma}_B$ on $\bold{GF}(q)$ are used at the two sources. The two
parity check matrices have the same size of $M \times N$ and the
same locations of non-zero's, where $N$ is the codeword length and $M$ is the number of parity check symbols. Let $\Gamma^A_{mn}$ and $\Gamma^B_{mn}$ denote
the elements at the $m$-th row and $n$-th column of $\bold{\Gamma}_A$ and
$\bold{\Gamma}_B$, respectively. $M_m=\{n:\Gamma_{mn}\neq 0\}$ denotes the set
of column locations of the non-zero's in the $m$-th row; $M_{m
\backslash n}=\{n':\Gamma_{mn'}\neq 0\}\backslash \{n\}$ denotes the set
of column locations of the non-zero's in the $m$-th row, excluding
location $n$. Likewise, $N_n=\{m:\Gamma_{mn}\neq 0\}$ and $N_{n \backslash
m}=\{m':\Gamma_{m'n}\neq 0\}\backslash \{m\}$ denotes the set of row
locations of the non-zero's in the $n$-th column and those excluding location
$m$. Note that in the following sections the arithmetic operations
are all in $\bold{GF}(q)$ unless specified otherwise.

\subsubsection{Check-relation-tab}\label{PCD_table_table}

Considering the encoding characteristics of $\bold{\Gamma}_A$ and
$\bold{\Gamma}_B$, we have the set of parity check equations, for each $m=1,\ldots,M$, as
follows:
\begin{equation}\label{eq:check_1}
\begin{array}{lll}
\sum\limits_{n \in M_m} \bold{C}_A(n) \times \Gamma^A_{mn} = 0,\\
\sum\limits_{n \in M_m} \bold{C}_B(n) \times \Gamma^B_{mn} = 0.
\end{array}
\end{equation}

For each set of the above check equations, we construct one check-relation-tab. The $m$-th check-relation-tab essentially characterizes the joint constraint
of the symbol pairs $\{\bold{C}_A(n), \bold{C}_B(n)\}$ at locations
$n \in M_m$. It consists of two parts,
one for virtual encoder, and the other for PCD decoder.

In virtual encoder, without loss of generality, we assume that the symbol pair
$\{\bold{C}_A(n), \bold{C}_B(n)\}$ at a random
location $n$ is an unknown parity check symbol pair while those $\{\bold{C}_A(n'), \bold{C}_B(n')\}$ at the other non-zero locations $n' \in M_{m \backslash n}$ are the known information symbol pairs. The parity check symbol pair and the information symbol pair are assumed as the symbol pairs composed of the unknown parity check symbols and the known information symbols respectively. We obtain the possible values for
$\{\bold{C}_A(n), \bold{C}_B(n)\}$ at the given $n$ base on
(\ref{eq:check_1}) through enumerating all values for
$\{\bold{C}_A(n'), \bold{C}_B(n')\}$ at locations $n' \in M_{m
\backslash n}$. Then, the symbol pairs $\{\bold{C}_A(n),
\bold{C}_B(n)\}$ at all locations $n \in M_m$ are mapped to
$\bold{C}_C(n)$, according to a given PLNC mapping rule $\bold{C}_C = {\cal M}(\bold{C}_A, \bold{C}_B)$. Since the number of symbol pairs $\{\bold{C}_A(n), \bold{C}_B(n)\}$
mapped to each element $\bold{C}_C(n)$ may not be the same, the
probability of occurrence for each element $\bold{C}_C(n)$ should be
computed separately.

For PCD decoder, without loss of generality, we assume that
the element $\bold{C}_C(n)$ at location $n$ is known. That is to
say, the set of symbol pairs mapped to $\bold{C}_C(n)$ is
given. Then, we should compute the probability of occurrence for
the corresponding possible values $\bold{C}_C(n')$ at locations $n'
\in M_{m \backslash n}$, named as weighted factor $F_W$, by classifying the aforementioned
probability of each element which is generated by the virtual
encoder.

In the following we shall use a toy example to demonstrate the detailed construction of the check-relation-tab as mentioned above. Consider the two source LDPC codes on $\bold{GF}(2)$ with
\[
\bold{\Gamma}_A = \bold{\Gamma}_B = \left[
\begin{array}{cccccccccccc}
1 & 0 & 1 & 0 & 0 & 1  \\
1 & 1 & 0 & 1 & 0 & 0  \\
0 & 0 & 1 & 1 & 1 & 0  \\
0 & 1 & 0 & 0 & 1 & 1
\end{array}
\right],
\]
where the code length is 6 and the row
weight and column weight are 3 and 2 respectively.

In order to indicate that our proposed PCD approach can deal with
any PLNC mapping under the exclusive law (\ref{eq:exclusive_law}),
a special non-linear clustering of the codeword pair, denoted as ${\cal M}_{nl}$, is considered here and characterized as:
$$\Big\{\bold{C}_A(n), \bold{C}_B(n)\Big\}\rightarrow\{\bold{C}_C(n)\}:$$
$$\Big\{(0,1)\Big\}\rightarrow\{a\}, \Big\{(0,0),(1,1)\Big\}\rightarrow\{b\},\Big\{(1,0)\Big\}\rightarrow\{c\}.$$

Each symbol pair $\Big\{\bold{C}_A(n), \bold{C}_B(n)\Big\}$ is mapped to one
of the three elements $\{a,b,c\}$. Here, to avoid confusion, we let $\{a,b,c\}$ indicate the decoding symbols (they are not necessarily equal to broadcasted symbols) during the period of operating PCD algorithm. Such mapping ${\cal M}_{nl}$ may be appropriate when the channel ratio satisfies $H_{BC}/H_{AC} \cong -1$ and the BPSK modulation is applied at two source nodes.
\begin{figure}
\centering
\includegraphics[width=3.4in]{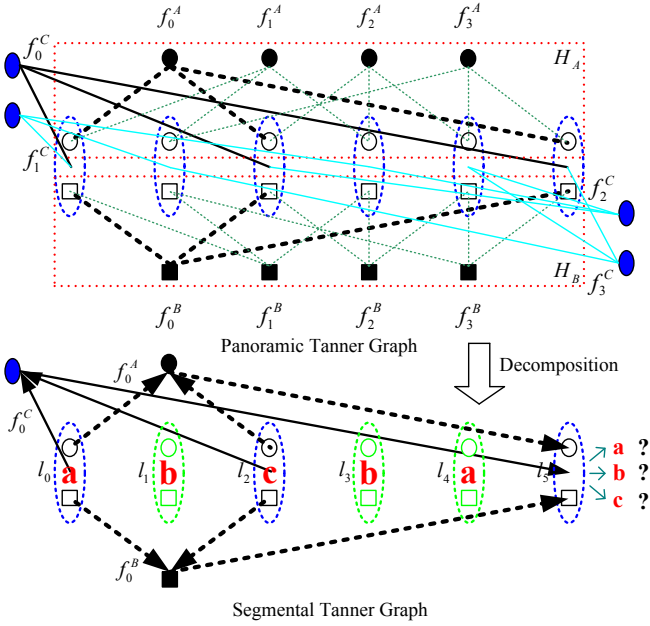}
\caption{The Tanner graph of the check functions versus codeword pair
at the relay.} \label{fig:tanner}
\end{figure}

The corresponding Tanner graph of the virtual LDPC code observed at the relay is shown in Fig.~\ref{fig:tanner}.
The solid circles and squares denote the check functions at each
source node, while non-solid circles and squares denote the
transmitted symbols of a code. In like manner, the solid and
non-solid ellipses denote the check functions and the corresponding
received symbol pairs. $f_m^s$ denotes the $m$-th check function of
the LDPC code at the node $s$, where $m \in [0,3], s \in \{A,B,C\}$.

Next, we will derive the check function of $f_0^C$ from $f_0^A$ and
$f_0^B$ using the segmental Tanner graph in Fig.~\ref{fig:tanner},
which is separated from the panoramic Tanner graph. $l_n$ denotes
the symbol pair $(\bold{C}_A(n), \bold{C}_B(n))$, for $n \in
[0,5]$. Then, the first set of constraint equations associated with the $0^{th}$ row of $\bold{\Gamma}_A$ and $\bold{\Gamma}_B$ can be written as:
\begin{equation}\label{eq:check_2}
\begin{array}{lll}
\bold{C}_A(0) + \bold{C}_A(2) + \bold{C}_A(5) = 0, \\
\bold{C}_B(0) + \bold{C}_B(2) + \bold{C}_B(5) = 0.
\end{array}
\end{equation}

Given the ${\cal M}_{nl}$ mapping, the so-called check-relation-tabs for virtual encoder and PCD decoder are
generated, as shown in Table \ref{tab:pcd_2_ary}. For virtual encoder, without loss of generality, we
assume that the symbol pairs $l_0$ and $l_2$ have
been known, e.g. $l_0 = a$ and $l_2 = c$, as shown in Fig.~\ref{fig:tanner}. The possible value for the parity check symbol pair $l_{5}$ can be obtained by (\ref{eq:check_2}). It is important to note that each element constrained by the pairwise
check functions $f_m^C$ corresponds to one symbol pair in
$\big\{l_n, n \in \{0,2,5\}\big\}$, the range of which is in
$\{a,b,c\}$.
\begin{table}[!h]
\tabcolsep 4.8pt \caption{Check-relation-tabs of $f_0^C$ for binary
LDPC codes with BPSK modulation}
\begin{center}\label{tab:pcd_2_ary}
\def\temptablewidth{0.49\textwidth}
{\rule{\temptablewidth}{1.5pt}}
\begin{tabular*}{\temptablewidth}{c|c||c|c||c|c|||c||c||c|c}
\multicolumn{6}{c|||} {Virtual Encoder} & \multicolumn{4}{c}{PCD
Decoder} \\
 \hline\hline
 \multicolumn{2}{c||} {$(0,0)$} & \multicolumn{2}{c||} {$(2,2)$} & \multicolumn{2}{c|||} {$(5,5)$} &$(5,5)$ & $F_W$ & $(0,0)$ & $(2,2)$
\\  \hline\hline
    a    &  (0,1)    &  a   &  (0,1)   &  1b  &  (0,0)   &  a &  0.5   &  a    &  b  \\  \hline
    \multirow{2}*{a}    &  \multirow{2}*{(0,1)}    &  \multirow{2}*{b}   &  (0,0)  & 0.5a  &  (0,1)  &  a &  0.5   &  b    &  a  \\
    \cline{7-10}
        & & & (1,1)& 0.5c & (1,0) & a &  0.5   &  b    &  c  \\  \hline
        a    &  (0,1)    &  c   &  (1,0)   &  1b  &  (1,1)   &  a &  0.5   &  c    &  b  \\  \hline
    \multirow{2}*{b}    &(0,0)    &  \multirow{2}*{a}   &  \multirow{2}*{(0,1)}  & 0.5a  &  (0,1)  &  b &  1   &  a    &  a  \\
    \cline{7-10}
       & (1,1) & & & 0.5c & (1,0) & b &  1   &  a    &  c  \\  \hline
     \multirow{2}*{b}    &(0,0)    &  \multirow{2}*{b}   &  (0,0)  & \multirow{2}*{1b}  &  (0,0)  &  b &  1   &  b    &  b  \\
    \cline{7-10}
       & (1,1) & &(1,1) &  & (1,1) & b &  1   &  c    &  a  \\  \hline
       \multirow{2}*{b}    &(0,0)    &  \multirow{2}*{c}   &  \multirow{2}*{(1,0)}  & 0.5a  &  (0,1)  &  b &  1   &  c    &  c  \\
    \cline{7-10}
       & (1,1) & & &  0.5c & (1,0) & c &  0.5   &  a    &  b  \\  \hline
       c    &  (1,0)    &  a   &  (0,1)   &  1b  &  (1,1)   &  c &  0.5   &  b    &  a  \\  \hline
       \multirow{2}*{c}    & \multirow{2}*{(1,0)}    &  \multirow{2}*{b}   &  (0,0)  & 0.5a  &  (0,1)  &  c &  0.5   &  b    &  c  \\
    \cline{7-10}
       &  & & (1,1)&  0.5c & (1,0) & c &  0.5   &  c    &  b  \\  \hline
      c    &  (1,0)    &  c   &  (1,0)   &  1b  &  (0,0)   &   &     &      &    \\  \hline
       \end{tabular*}
       {\rule{\temptablewidth}{1.5pt}}
       \end{center}
       \end{table}

Through reverse derivation, the check-relation-tab for PCD decoder are given.
Therein, $F_W$ is the
weighted factor which should be multiplied with the probability of
the two elements behind. Then, we just need to compute the
probability of occurrence of the corresponding possible values
$\{a,b,c\}$ for $l_0$ and $l_2$ one by one. Interestingly,
using the binary LDPC codes or the $q$-ary $(q
> 2)$ codes with the non-zero elements $\{\eta,\ldots,\eta \in {\cal
Z}_{q}\}$, we have the same probability distribution of occurrence
among $\big\{l_n, n \in \{0,2,5\}\big\}$ for a given PLNC
mapping, like $f_m^s = f_{m'}^s, m \neq m'$.

\subsection{Pairwise check decoding (PCD) algorithm}

After obtaining the check-relation-tabs, the PCD algorithm can be
readily carried out. Note that the locations of non-zeros in the
$q$-ary LDPC codes used at two source nodes are still valid here. We
only change the check functions of symbol pairs from
$\{f_m^A,f_m^B\}$ to $f_m^C$ by the derived check-relation-tab. Define
$u_{nm}^k = Pr(\bold{C}_C(n)= k|\bold{Y}_C(n),w_{mn})$; $v_{mn}^k = Pr(f_m^C
\text{satisfied}|\bold{C}_C(n)= k,t_{nm})$. Let $t_{nm}$ denotes the messages to
be passed from symbol node $\bold{C}_C(n)$ to check node $f_m^C$;
$w_{mn}$ denotes the messages to be
passed from check node $f_m^C$ to symbol node $\bold{C}_C(n)$.
Suppose that $\{m_k\}, k \in {\cal Z}_{q'}$, denote the index-set of
rows, which have the same target value in the check-relation-tab for PCD decoder. Therein, the element
at $(n,n)$ is generated as $\{k\}$ and $m_{tab}$ means the index of
row.

For a given PLNC mapping ${\cal M}$, we compute for
each $[m,n]$ that satisfies $\Gamma_{mn} \neq 0$.

1. Initialization

Compute the initial value of each $u_{nm}^{k}$ and each $t_{nm}$ as:
\begin{equation}
\label{eq:pcd_1}
\begin{array}{lll}
u_{nm}^{'k} \\= \sum\limits_{(\bold{C}_A(n), \bold{C}_B(n)):
\bold{C}_C(n)=k} {Pr\Big((\bold{C}_A(n),
\bold{C}_B(n))|\bold{Y}_C(n)\Big)}; \\
t_{nm} =(u_{nm}^k,k \in {\cal Z}_{q'}) = \Big((1/\sum\limits_{k\in {\cal Z}_{q'}}u_{nm}^{'k}) u_{nm}^{'k},k \in {\cal Z}_{q'}\Big),
\end{array}
\end{equation}
where $Pr\Big((\bold{C}_A(n), \bold{C}_B(n))|\bold{Y}_C(n)\Big)$ is the
probability of $(\bold{C}_A(n),\bold{C}_B(n))$ given $\bold{Y}_C(n)$
is received. Obviously, we have $u_{nm_1}^k=u_{nm_2}^k$ even if $m_1 \neq m_2, m_1,m_2 \in N_{n}.$ For example, given ${\cal M}_{nl}$ mapping\footnote{Note that the definition of ${\cal M}_{nl}$ mapping, which will be used as an example in this subsection, is similar to the definition described in Subsection~\ref{PCD_table_table}.}, we have
\begin{equation}
\begin{array}{lll}
u_{nm}^{'a}&=&\exp\Big(\frac{-(\bold{Y}_C(n)-(H_{AC}-H_{BC}))^2}{2\sigma_C^2}\Big), \\ u_{nm}^{'b}&=&\exp\Big(\frac{-(\bold{Y}_C(n)-(H_{AC}+H_{BC}))^2}{2\sigma_C^2}\Big) \\&+& \exp\Big(\frac{-(\bold{Y}_C(n)-(-H_{AC}-H_{BC}))^2}{2\sigma_C^2}\Big), \\ u_{nm}^{'c}&=&\exp\Big(\frac{-(\bold{Y}_C(n)-(-H_{AC}+H_{BC}))^2}{2\sigma_C^2}\Big).
\end{array}
\end{equation}

2. First half round iteration: from symbol node $\bold{C}_C(n)$ to check node $f_m^C$
\begin{equation}
\label{eq:pcd_2}
\begin{array}{lll}
v_{mn}^k = \sum\limits_{m_{tab} \in \{m_k\}} F_W(m_{tab})  \prod\limits_{n'\in M_{m \backslash n}}t_{n'm}(m_{tab});\\
w_{mn}=(v_{mn}^k,k \in {\cal Z}_{q'}),
\end{array}
\end{equation}
where $t_{n'm}(m_{tab})$ denotes that $k$ of $(u_{n'm}^k,k \in {\cal
Z}_{q'})$ is the designated elements at the $m_{tab}$-th row in the check-relation-tab for PCD decoder. For ${\cal M}_{nl}$ mapping, we have
\begin{equation}
\begin{array}{lll}
v_{05}^a=0.5(u_{00}^a u_{20}^b + u_{00}^b u_{20}^a + u_{00}^b u_{20}^c + u_{00}^c u_{20}^b),\\
v_{05}^b=u_{00}^a u_{20}^a + u_{00}^a u_{20}^c + u_{00}^b u_{20}^b + u_{00}^c u_{20}^a + u_{00}^c u_{20}^c,\\
v_{05}^c=0.5(u_{00}^a u_{20}^b + u_{00}^b u_{20}^a + u_{00}^b u_{20}^c + u_{00}^c u_{20}^b).
\end{array}
\end{equation}

3. Second half round iteration: from check node $f_m^C$ and initial value to symbol node $\bold{C}_C(n)$
\begin{equation}
\label{eq:pcd_3}
\begin{array}{lll}
u_{nm}^{'k} = p_k^{-(o_n-1)} u_{nm}^k  \prod\limits_{m'\in N_{n \backslash m}}w_{m'n};\\
t_{nm}=(u_{nm}^k,k \in {\cal Z}_{q'})= \Big((1/\sum\limits_{k\in {\cal Z}_{q'}}u_{nm}^{'k}) u_{nm}^{'k},k \in {\cal Z}_{q'}\Big),
\end{array}
\end{equation}
where $p_k$ denotes the average probability of occurrence of element
$k$ and $o_n$ indicates column weight for the $n$-th symbol node.
Let $P_n=\{p_k, k \in {\cal Z}_{q'}\}$, e.g., we have
$P_n=\{\frac{1}{4},\frac{1}{2},\frac{1}{4}\}$ for
${\cal M}_{nl}$ mapping. Note that in (\ref{eq:pcd_3}) the firstly mentioned $u_{nm}^k$ denotes the corresponding value generated in Step 1. Here it operates as extra information and takes part in the Tanner graph.

4. Soft decision
\begin{equation}
\label{eq:pcd_4}
\begin{array}{lll}
U_n^{'k} = p_k^{-o_n} u_{nm}^k  \prod\limits_{m \in N_{n}}w_{mn};\\
T_n=(U_n^k,k \in {\cal Z}_{q'})
= \Big((1/\sum\limits_{k\in {\cal Z}_{q'}}U_n^{'k}) U_n^{'k},k \in {\cal Z}_{q'}\Big).
\end{array}
\end{equation}
Here $u_{nm}^k$ also denotes the corresponding value generated in Step 1.

5. Hard decision
\begin{equation}
\label{eq:pcd_5} \bold{\hat {\bold{C}}}_C(n) = {\rm{arg}}\max_{{\cal
M'}:\{k'\}} \sum \limits_{\stackrel{{\cal M}\rightarrow {\cal M}'}{{\cal M}:\{k\},{\cal M'}:\{k'\}}} U_n^k.
\end{equation}

If $\bold{C}_C$ satisfies the applied ${\cal M'}$ mapping's
check-relation-tab for virtual encoder or the number of iterations
exceeds a certain value, then the algorithm stops, otherwise we go
to Step 2. Note that ${\cal M'}$ mapping denotes the second mapping used in hard decision, which may be different from the ${\cal M}$ mapping applied in soft iterations. Fortunately, the cardinality of the ${\cal M'}$ mapping is always far less than that of the ${\cal M}$ mapping. That is to say, the satisfaction of the check-relation-tabs will be checked more effectively. For ${\cal M}_{nl}$ mapping, ${\cal M'}_{nl}$ mapping is selected as $\{b\}\rightarrow\{0\}$ and $\{a,c\}\rightarrow\{1\}$. Namely, $\{k\}=\{a,b,c\}$ and $\{k'\}=\{0,1\}$. Therefore, $P(0)=P(b)$ and $P(1)=P(a)+P(c)$ should be carried out in hard decision. Therein, $P(i)$ denotes the probability of occurrence of the element $i$, and \{0,1\} indicate the PLNC mapped symbols which will be transmitted in BC phase.

So far, the whole PCD algorithm for arbitrary PLNC
mapping is presented.

\subsection{Convergence behavior}

There are many methods that can be used to investigate the
convergence behavior of iterative decoding. Examples are the density
evolution algorithm~\cite{IT01:Richardson_f} and the extrinsic
information transfer (EXIT) chart~\cite{TCOM01:Brink}, both of which
are suitable for LDPC codes over Gaussian channels. However, the
virtual LDPC code, namely, check-relation-tab in the considered TWR
model is different from conventional LDPC codes. Moreover, the
proposed check-relation-tab is determined by the selected PLNC
mapping. That is to say, several check-relation-tabs should be
selected adaptively at the relay. To the best of our knowledge,
there is no any literature to solve the analogous problem up to now.
Similar to~\cite{JSAC09:Zhang}, we resort to simulations in Section
VII for confirming the convergence behavior of the proposed PCD
algorithm.

\subsection{Complexity analysis}

Since decoding complexity of the proposed PCD approach is basically
determined by the generated check-relation-tabs, we focus complexity
calculation on the number and size of the check-relation-tabs. Note that the proposed check-relation-tab is generated by certain two correlative rows with the same index of non-zero elements. The two correlative rows is defined as a certain two-rows, in which each row has the same row index of the respective LDPC code, e.g., two $0^{th}$ rows of $\bold{\Gamma}_A$ and
$\bold{\Gamma}_B$. Let $M \times N$, $d_r$ and $r_\kappa$ denote matrix size, maximum
row weight and the $\kappa$-th row weight in degree
distributions of an arbitrary irregular LDPC code, respectively. The
number of the check-relation-tabs for a given PLNC mapping, denoted
as $N_T$, is upper bounded by
\begin{equation}\label{eq:complexity_number}
N_T \leq \sum\limits_{r_\kappa = 2}^{d_r}
\min\Big(\lambda_{r_\kappa} M, q^{2(r_\kappa - 1)}\Big)r_\kappa,
\end{equation}
where $\lambda_{r_\kappa}$ is a ratio of the number of rows with row
weight $r_\kappa$ to $M$. The equality can be reached when the non-zero
elements of both two correlative rows are different from each other.

We also obtain the size of the check-relation-tab for virtual encoder with a given PLNC mapping as
\begin{equation}\label{eq:complexity_size_encoder}
S_E = {q'}^{(r_\kappa - 1)},
\end{equation}
where $q'$ denotes the range of PLNC mapped symbols at the relay, $q
\leq q' \leq q^2$.

Furthermore, the size range of the
check-relation-tab for PCD decoder is obtained as
\begin{equation}\label{eq:complexity_size_decoder}
{q'}^{(r_\kappa - 1)} \leq S_D \leq {q'}^{r_\kappa}.
\end{equation}

The number and size of the check-relation-tabs are determined by the selected PLNC mapping and the applied
optimizations. We will state it in Section~\ref{TS-CNC-PCD} in detail.

\section{Two-stage CNC mapping with PCD}\label{TS-CNC-PCD}

Recall that the design criterion of ${\cal M}(\bold{C}_A,
\bold{C}_B)$ is to maximize the MED, as stated in Section IV.
However, it is difficult to directly maximize the MED. In this
section, we shall propose a two-stage CNC mapping based on PCD decoding, which optimizes the symbol distance first and then the Hamming distance.

\subsection{Two-stage CNC mapping with PCD (TS-CNC-PCD)}

In this method, the traditional CNC mapping proposed
in~\cite{JSAC09:Koike-Akino} is divided to two steps by a maximum splitting and minimum merging (MSMM) strategy for maximizing
the symbol distance first and then a minimum correlation optimization (MCO) method based on the proposed PCD
approach is presented to optimize the Hamming distance.

\subsubsection{Channel coding structure at the relay}
\begin{figure*}
\centering
\includegraphics[width=4.4in]{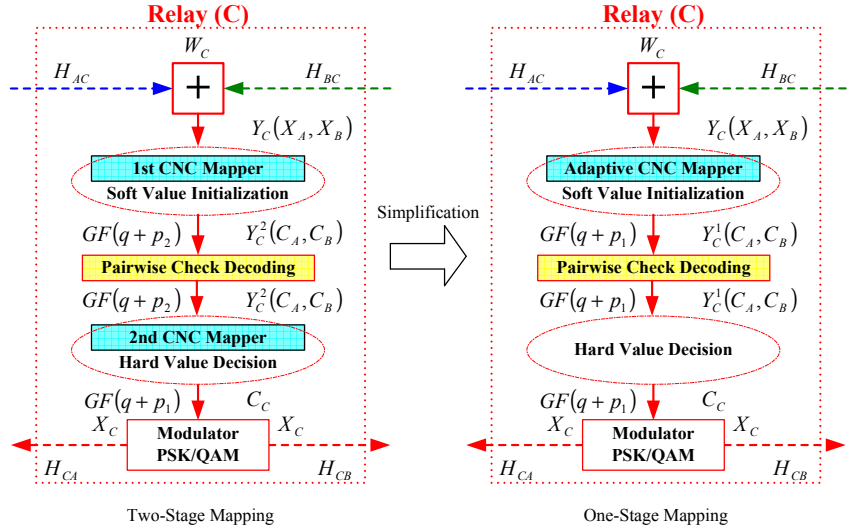}
\caption{Partial decoding model at the relay node for TWR fading
channels.} \label{fig:partial_decoding}
\end{figure*}

The proposed TS-CNC-PCD is composed of the first CNC mapper
($\bold{GF}(q+p_2)$), PCD decoder ($\bold{GF}(q+p_2)$), the second
CNC mapper ($\bold{GF}(q+p_1)$) and $(q+p_1)$PSK/QAM modulator as
shown in Fig.~\ref{fig:partial_decoding}. Upon receiving the
superimposed signal, the relay initializes the soft value for PCD
decoder by taking into account the 1st CNC mapping. The output of
the PCD decoder, i.e., the 2nd CNC mapped codeword packet,
$\bold{C}_C = {\cal M}(\bold{C}_A, \bold{C}_B)$, is then modulated to
$(q+p_1)$PSK/QAM accordingly to obtain $\bold{X}_C$. The cardinality of the received symbol pairs at the relay node is $q^2$. With the first application of the PLNC which is denoted as the 1st CNC mapping, we reduce the cardinality to $q+p_2$ by clustering some received symbol pairs into one decoding symbol. Next, we make further reduction in the cardinality through the second application of the PLNC, i.e., the 2nd CNC mapping. Wherein, we cluster some decoding symbols into one broadcasted symbol. Here, both $p_1$ and $p_2$, $p_1,p_2 \in {\cal Z}_{q^2-q}$, represent the possible expanding
of the cardinality compared to $q$. For example, we have $p_1 = 0(1)$, $p_2=5(8)$ for the QPSK(5QAM) at the relay node
when QPSK ($q=4$) is used at two sources. Note that the 1st and 2nd CNC mappings denote the symbol pair mappings while the aforementioned ${\cal M}$ mapping indicates the codeword pair mapping.

\subsubsection{Two-stage CNC mapping}

Since the best CNC mappings for any choice of constellation size have been presented in~\cite{JSAC09:Koike-Akino}, we just use for reference and generate the two-stage CNC mapping. Replacing ${\cal Z}_4 \times {\cal Z}_4$ by ${\cal Z}_q \times {\cal Z}_q$, we can easily obtain the best CNC mappings for $q$-ary modulations using the Algorithm 1 in~\cite{JSAC09:Koike-Akino}. We enlarge the MED between two distinct codewords (e.g. $\bold{C}_C^l$ and
$\bold{C}_{C}^{l'}$) by extending the symbol distance between two individual coded symbols (namely CNC mapped clusters), if the codewords are considered instead of uncoded symbols. Using the proposed MSMM strategy, we divide the traditional CNC mapping (denoted as ${\cal M}^t$) into
the 1st CNC mapping (denoted as ${\cal M}^s$) and the 2nd CNC mapping (denoted as ${\cal M}^h$) for soft
value initialization and hard value decision, respectively. For certain $H_{AC}$ and $H_{BC}$, the proposed MSMM strategy is formulated as

a) Maximum splitting (${\cal M}^t \rightarrow {\cal M}^s$): Any two symbol pairs, belonged to a ${\cal M}^t$ mapped cluster, should be split into two ${\cal M}^s$ mapped clusters, if the distance between which is greater than $d_{max}$. All ${\cal M}^s$ mapped clusters, split from an identical ${\cal M}^t$ mapping, generate a ${\cal M}^s$ mapping. Note that two distinct ${\cal M}^t$ mappings can generate an identical ${\cal M}^s$ mapping for the same channel condition, while an identical ${\cal M}^t$ mapping can be divided into more than one distinct ${\cal M}^s$ mappings for different channel conditions.

b) Minimum merging (${\cal M}^s \rightarrow {\cal M}^h$): Any two distinct ${\cal M}^s$ mapped clusters, split from a ${\cal M}^t$ mapped cluster, should be merged into a ${\cal M}^h$ mapped cluster. All ${\cal M}^h$ mapped clusters, merged by an identical ${\cal M}^s$ mapping, generate a ${\cal M}^h$ mapping. Any two distinct ${\cal M}^h$ mappings also should be merged into one ${\cal M}^h$ mapping if the minimum distances of which are both equal to $d_{min}$.

Therein, $d_{max}$ and $d_{min}$ denote the maximum and minimum values among whole distances between any two ${\cal M}^t$ mapped clusters, respectively. A mapping is composed by several clusters. Several symbol pairs form a cluster.
\begin{figure}
\center{\subfigure[1st CNC Mapping (Soft Value
Initialization)]{\includegraphics[width=2.4in]{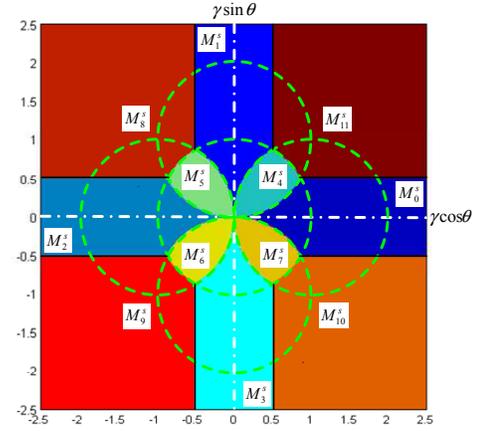}
\label{a}} \hfil \subfigure[2nd CNC Mapping (Hard Value
Decision)]{\includegraphics[width=2.4in]{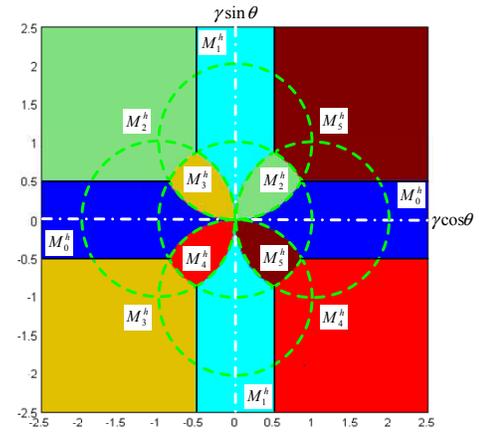}
\label{b}}} \caption{Two-stage CNC mapping according to the channel
ratio $H_{BC}/H_{AC}=\gamma(\cos\theta+j\sin\theta)$
when $q = 2^2$.} \label{fig:pcd_sel}
\end{figure}
\begin{table*}[t]
\tabcolsep 4.5pt \caption{Two-stage CNC mappings for 4-ary LDPC
codes with QPSK modulation}
\begin{tiny}
\begin{center}\label{tab:TS_CNC}
\def\temptablewidth{0.9\textwidth}
{\rule{\temptablewidth}{1.5pt}}
\begin{tabular*}{\temptablewidth}{c||cccc|cccc|cccc|cccc||c}
 & $(0,0)$ & $(0,1)$& $(0,2)$& $(0,3)$& $(1,0)$& $(1,1)$& $(1,2)$& $(1,3)$& $(2,0)$& $(2,1)$& $(2,2)$& $(2,3)$& $(3,0)$& $(3,1)$& $(3,2)$&
 $(3,3)$ & Cardinality
\\  \hline\hline
${\cal M}_0^s$ & $f$ & $b$ & $c$ & $a$ & $b$ & $g$ & $a$ & $d$ & $c$ & $a$ & $h$ & $e$ & $a$ & $d$ & $e$ & $i$ & 9 \\
${\cal M}_2^s$ & $a$ & $b$ & $c$ & $f$ & $d$ & $a$ & $g$ & $c$ & $e$ & $h$ & $a$ & $b$ & $i$ & $e$ & $d$ & $a$ & 9  \\  \hline ${\cal
M}_0^h$ & $a'$ & $b'$ & $c'$ & $d'$  & $b'$    &  $a'$   & $d'$   & $c'$ & $c'$ & $d'$   & $a'$ & $b'$ & $d'$ & $c'$   &  $b'$   & $a'$  & 4  \\ \hline\hline
${\cal M}_1^s$ & $b$ & $a$ & $f$ & $c$ & $g$ & $d$ & $b$ & $a$ & $a$ & $e$ & $c$ & $h$ & $d$ & $i$ & $a$ & $e$ & 9 \\
${\cal M}_3^s$ & $b$ & $f$ & $a$ & $c$ & $a$ & $c$ & $d$ & $g$ & $h$ & $b$ & $e$ & $a$ & $e$ & $a$ & $i$ & $d$ & 9  \\  \hline ${\cal
M}_1^h$ & $a'$ & $b'$ & $c'$ & $d'$  & $c'$    &  $d'$   & $a'$   & $b'$ & $b'$ & $a'$   & $d'$ & $c'$ & $d'$ & $c'$   &  $b'$   & $a'$  & 4  \\ \hline\hline
${\cal M}_4^s$ & $e$ & $a$ & $f$ & $b$ & $g$ & $h$ & $a$ & $c$ & $b$ & $d$ & $i$ & $j$ & $c$ & $k$ & $d$ & $l$ & 12 \\
${\cal M}_8^s$ & $a$ & $b$ & $e$ & $f$ & $g$ & $c$ & $h$ & $d$ & $c$ & $i$ & $d$ & $j$ & $k$ & $l$ & $a$ & $b$ & 12 \\  \hline ${\cal
M}_2^h$ & $b'$ & $c'$ & $a'$ & $d'$  &  $a'$    &  $d'$   &  $c'$ & $e'$ & $d'$ & $b'$ & $e'$ & $a'$ & $e'$    &  $a'$   &  $b'$   &  $c'$  & 5 \\ \hline\hline
${\cal M}_5^s$ & $a$ & $b$ & $e$ & $f$ & $g$ & $c$ & $h$ & $a$ & $d$ & $i$ & $b$ & $j$ & $k$ & $l$ & $c$ & $d$  & 12 \\
${\cal M}_9^s$ & $a$ & $e$ & $b$ & $f$ & $c$ & $d$ & $g$ & $h$ & $i$ & $j$ & $c$ & $d$ & $k$ & $a$ & $l$ & $b$ & 12 \\  \hline ${\cal
M}_3^h$ & $b'$ & $c'$ & $d'$ & $a'$  &  $c'$    &  $e'$   &  $a'$ & $b'$ & $d'$ & $a'$ & $c'$ & $e'$ & $a'$    &  $b'$   &  $e'$   & $d'$  & 5 \\  \hline\hline
${\cal M}_6^s$ & $a$ & $e$ & $b$ & $f$ & $c$ & $b$ & $g$ & $h$ & $i$ & $j$ & $d$ & $a$ & $k$ & $d$ & $l$ & $c$ & 12 \\
${\cal M}_{10}^s$ & $e$ & $f$ & $a$ & $b$ & $c$ & $g$ & $d$ & $h$ & $i$ & $c$ & $j$ & $d$ & $a$ & $b$ & $k$ & $l$ & 12 \\  \hline ${\cal
M}_4^h$ & $b'$ & $a'$ & $c'$   &  $d'$  &  $e'$    &  $c'$   & $b'$ & $a'$ & $a'$ & $e'$ & $d'$ & $b'$ &  $c'$    &  $d'$   &  $a'$   & $e'$ & 5 \\  \hline\hline
${\cal M}_7^s$ & $e$ & $f$ & $a$ & $b$ & $b$ & $g$ & $c$ & $h$ & $i$ & $a$ & $j$ & $d$ & $d$ & $c$ & $k$ & $l$ & 12 \\
${\cal M}_{11}^s$ & $e$ & $a$ & $f$ & $b$ & $g$ & $h$ & $c$ & $d$ & $c$ & $d$ & $i$ & $j$ & $a$ & $k$ & $b$ & $l$ & 12 \\  \hline ${\cal
M}_5^h$ & $a'$ & $b'$ & $c'$ &  $d'$  &  $d'$    &  $a'$   &  $e'$ & $c'$ & $e'$ & $c'$ & $a'$ & $b'$ &  $b'$    &  $e'$   &  $d'$   & $a'$ & 5 \\  \hline
       \end{tabular*}
       {\rule{\temptablewidth}{1.5pt}}
       \end{center}
       \end{tiny}
       \end{table*}
\begin{figure*}[t]
\centering
\includegraphics[width=5.2in]{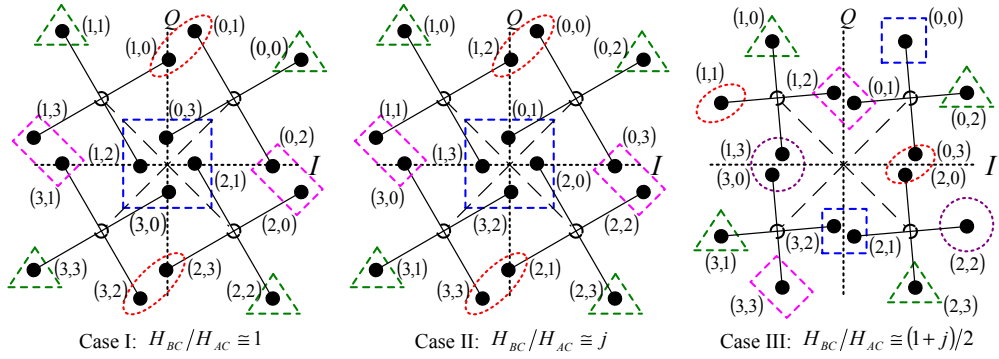}
\caption{Received signal constellation with two-stage CNC mapping at the
relay.} \label{fig:received_constellation}
\end{figure*}

For instance, we consider the 4-ary LDPC codes with the QPSK modulation. The corresponding best CNC mappings have been presented in the Table I in~\cite{JSAC09:Koike-Akino}. As depicted in Fig.~\ref{fig:pcd_sel} and Table~\ref{tab:TS_CNC}, we generate 12 1st CNC mappings and 6 2nd CNC mappings, denoted as ${\cal M}_i^s, i \in [0,11]$ and ${\cal M}_j^h, j \in [0,5]$. Take
Fig.~\ref{fig:received_constellation}-Case I for example, we have four clusters according to the Table I in~\cite{JSAC09:Koike-Akino}, e.g., symbol pairs $\{(0,1),(1,0),(2,3),(3,2)\}$ should be clustered together. However, in two-stage CNC mapping, we only group together some symbol pairs (e.g. (0,1) and (1,0) in ${\cal
M}_0^s$) due to that the distance between which is much smaller than $d_{max}$. Whereas, we classify the other
symbol pairs (e.g. (2,3) and (3,2)), which are more far away from
(0,1) or (1,0), into another independent cluster. It is because that the distance between these two clusters \{(2,3),(3,2)\} and \{(0,1),(1,0)\} is greater than $d_{max}$. Certainly, these
separated clusters (e.g. (0,1),(1,0),(2,3) and (3,2)) are merged to
one cluster again in ${\cal M}_0^h$. Similarly, ${\cal
M}_1^s$/${\cal M}_1^h$ and ${\cal M}_4^s$/${\cal M}_2^h$ are
generated according to the proposed MSMM strategy as depicted in Fig.~\ref{fig:received_constellation}-Cases
II and III. Here, to
avoid confusion, we let $\{a',b',c',\dots \}$ indicate the
broadcasted symbols $\{0,1,2,\dots \}$, which are not same as the
decoding symbols $\{a,b,c,\dots \}$ during the period of operating
PCD algorithm.

\subsubsection{Check-relation-tabs and TS-CNC-PCD}

Note that only the 1st CNC mappings are operated before soft iteration in the proposed PCD approach, we
just need to generate the check-relation-tabs according to ${\cal
M}^s$. However, the additional check-relation-tabs for ${\cal
M}^h$ also need to be generated if the satisfaction of the check-relation-tabs should be checked during each iteration. Similar to Subsection V-A and V-B, we can obtain the check-relation-tabs for all 1st and 2nd CNC mappings easily and operate the proposed PCD algorithm directly, although we should modify the cardinality of the designed PLNC mapping here compared with that of Section V.

For instance, we have ${\cal M}_i^s, i \in [0,11]$ if the 4-ary LDPC codes and the QPSK modulation are applied at two source nodes. Take the ${\cal M}_4^s$ mapping of
Table~\ref{tab:TS_CNC} for example, which is displayed in
Fig.~\ref{fig:received_constellation}-Case III. Each symbol pair
$(C_A(n), C_B(n))$ is mapped to one of 12 elements based on
the fading conditions $H_{BC}/H_{AC} \simeq (1+j)/2$. All the
possible mappings are listed below:$$\{(0,1),(1,2)\}\rightarrow\{a\},
\{(0,3),(2,0)\} \rightarrow \{b\},$$ $$\{(1,3),(3,0)\} \rightarrow \{c\}, \{(2,1),(3,2)\} \rightarrow \{d\},$$ $$
\{(0,0)\}\rightarrow \{e\}, \{(0,2)\}\rightarrow \{f\}, \{(1,0)\}\rightarrow\{g\},$$ $$\{(1,1)\}\rightarrow\{h\},
\{(2,2)\}\rightarrow\{i\}, \{(2,3)\}\rightarrow\{j\},$$ $$\{(3,1)\}\rightarrow\{k\}, \{(3,3)\}\rightarrow\{l\}.$$
Each element constrained by the check functions $f_m^C$ corresponds
to one symbol pair in $\big\{(C_A(n),C_B(n)), n \in
M_m\big\}$, the range of which is in
$\{a,b,c,d,e,f,g,h,i,j,k,l\}$. According to the 2nd CNC mapping ${\cal M}_2^h$, we should operate $P(a')=P(f)+P(g)+P(j)+P(k),P(b')=P(e)+P(d),P(c')=P(a)+P(l),P(d')=P(b)+P(h),P(e')=P(c)+P(i)$ in hard decision.

\subsection{Minimum correlation optimization (MCO)}

Due to the irregular cardinality of the 1st CNC mappings, the size of mostly generated
check-relation-tabs for PCD decoder can approach the maximum value,
i.e., $S_D=q'^{r_\kappa}$. That is to say, we can not obtain larger
coding gains from these check-relation-tabs. Note that the
check-relation-tabs are fixed once a PLNC mapping and the
correlative rows of two LDPC codes ($\bold{\Gamma}_A$ and $\bold{\Gamma}_B$)
are all selected. Therefore, a kind of correlative rows
optimization, named as the minimum correlation optimization (MCO),
is proposed for selecting the better check-relation-tabs.
\begin{algorithm}
\KwData{given $r_\kappa$, $q$, $q'$, ${\cal M}$}
\KwResult{a collection of non-zero elements distribution for the
correlative rows} Initialization: $\{\eta_1, \eta_2, \dots,
\eta_{r_\kappa}\}=\{1,1, \dots, 1\}$, $\{\xi_1, \xi_2, \dots,
\xi_{r_\kappa}\}=\{1,1, \dots, 1\}$ \; Set the expected collection:
$C_{exp} = \O$\; Set the maximum average number of possible
generation: $G_{max} = q'$\; \While{$\eta_1 < q$} { $\forall {\cal
M}_i \in {\cal M}$, generate the relevant check-relation tabs\; Let
the temporary value $G_{{max}_{temp}}$ equal to the maximum average
number of possible generations of the lastly check-relation tabs\;
\If{$G_{{max}_{temp}} \leq G_{max}$} { \eIf{$G_{{max}_{temp}} =
G_{max}$}{ Add $\{\eta_1, \eta_2, \dots,
\eta_{r_\kappa}\}$,$\{\xi_1, \xi_2, \dots, \xi_{r_\kappa}\}$
$\rightarrow$ $C_{exp}$ \; }{Set $G_{max} = G_{{max}_{temp}}$,
$C_{exp}=\O$, add $\{\eta_1, \eta_2, \dots,
\eta_{r_\kappa}\}$,$\{\xi_1, \xi_2, \dots, \xi_{r_\kappa}\}$
$\rightarrow$ $C_{exp}$\;} } \eIf {$\xi_n(\eta_{n}) = q$, $n \in [2,
r_\kappa]$}{$\xi_n(\eta_{n}) = 1$, $\xi_{n-1}(\eta_{n-1}) =
\xi_{n-1}(\eta_{n-1}) +
1$\;}{$\xi_{r_\kappa}=\xi_{r_\kappa}+1$\;}\If{$\xi_1=q$}{$\{\xi_1,
\xi_2, \dots, \xi_{r_\kappa}\}=\{1,1, \dots, 1\}$,
$\eta_{r_\kappa}=\eta_{r_\kappa}+1$\;}} \caption{Minimum correlation
optimization (MCO)}\label{algorithm:MCO}
\end{algorithm}

Using the proposed MCO method (Algorithm~\ref{algorithm:MCO}), we
obtain a collection of non-zero elements distributions for certain correlative rows of $\bold{\Gamma}_A$ and $\bold{\Gamma}_B$. Among these distributions, some special correlative rows like $\{\eta,\dots,\eta\}, \eta \in {\cal Z}_{q}$, can generate the check-relation-tabs with a relatively small size. A check-relation-tab with a smaller size always has a lager Hamming distance. Certainly, larger
coding gains may be obtained by these check-relation-tabs. Moreover, these special correlative rows can also reduce the number of the check-relation-tabs sharply. In theory, it is
likely to generate different check-relation-tabs when the row weight
or the non-zero elements of any two rows of each LDPC code are not similar to each other. However,
only one check-relation-tab for the virtual encoder and the corresponding one check-relation-tab for PCD decoder needs to be obtained for
one or more selected adaptive PLNC mapping when the regular LDPC
codes are applied and the non-zero elements in every row follows
some special pattens like $\{\eta,\dots,\eta\}, \eta \in {\cal
Z}_{q}$. Considering the
trade-off of complexity and performance, we select the special correlative rows like $\{\eta,\dots,\eta\}, \eta \in {\cal Z}_{q}$.

\subsection{Simplification of TS-CNC-PCD}

To simplify the decoding process, the two-stage CNC mapping can be integrated into
one-stage mapping. Specifically, the identical mapping is used for
soft value initialization and hard value decision. As shown in
Fig.~\ref{fig:partial_decoding}, the simplification of TS-CNC-PCD is
composed of an adaptive CNC mapper ($\bold{GF}(q+p_1)$), PCD decoder
($\bold{GF}(q+p_1)$) and $(q+p_1)$PSK/QAM modulator. According to
the same design principle, we can generate the corresponding
check-relation-tabs. The expected error performance may be worse.
However, the complexity (size of check-relation-tab) is decreased
dramatically. More details can be found in~\cite{ICC10:We,ICC11:We}.

\section{Simulation results}

In this section, we present some simulation results to illustrate
the convergence behaviors and the coding gains of the proposed PCD
approach. For simplicity, each node uses the same transmission power
of one unit and observes the same noise power given by $\sigma^2$. Define an average
SNR per information symbol as $\frac{1}{2r\sigma^2}$, where $r$ is
the channel code rate. According to the proposed MCO method, we
generate a 4-ary code from a binary LDPC code "252.252.3.252", which
is produced by Mackay~\cite{URL09:Mackay}, through replacing
$\{1,\dots,1\}$ by $\{\eta,\dots,\eta\}$, $\eta \in {\cal Z}_4$.
Code length, code rate, row weight and column weight are 504, 0.5, 6
and 3, respectively.

The proposed partial decoding method, TS-CNC-PCD, is simulated in
kinds of TWR channels. The selections for the two-stage CNC mapping
and traditional CNC mapping are based on instantaneous realizations
of the channel gain pairs $\{H_{AC},H_{BC}\}$
using Fig.~\ref{fig:pcd_sel} (a,b) and Fig.~\ref{fig:pcd_sel} (b),
respectively.

For comparison, two benchmark systems are considered. One is the
uncoded case, where QPSK modulation is applied and the relay
demodulates using different types of PLNC mappings. The other is the coded
conventional XOR case, where the same 4-ary LDPC is applied at the
source and the relay performs traditional BP decoding with
conventional XOR mapping.

\subsection{Deterministic channels}
\begin{figure}
\center{\subfigure[$H_{AC}=H_{BC} =
1$]{\includegraphics[width=3.4in]{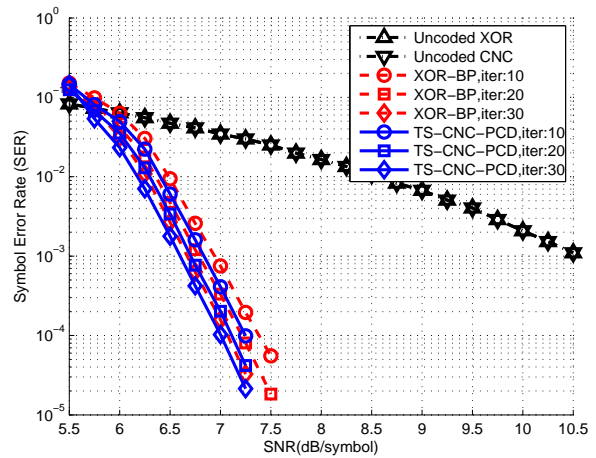} \label{a}}
\hfil \subfigure[$H_{AC}=1,H_{BC} =
j$]{\includegraphics[width=3.4in]{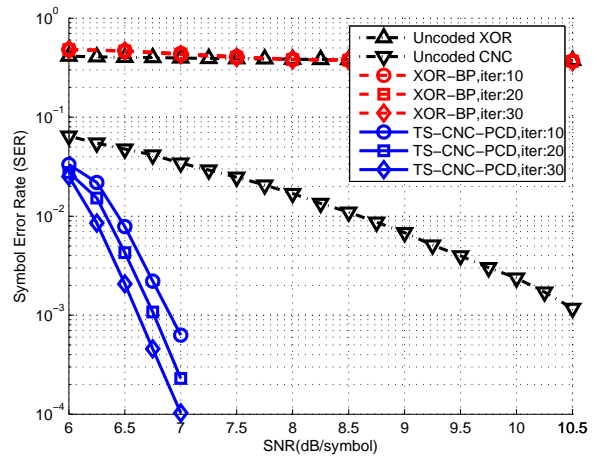} \label{b}}
\hfil \subfigure[$H_{AC}=1,H_{BC} =
(1+j)/2$]{\includegraphics[width=3.4in]{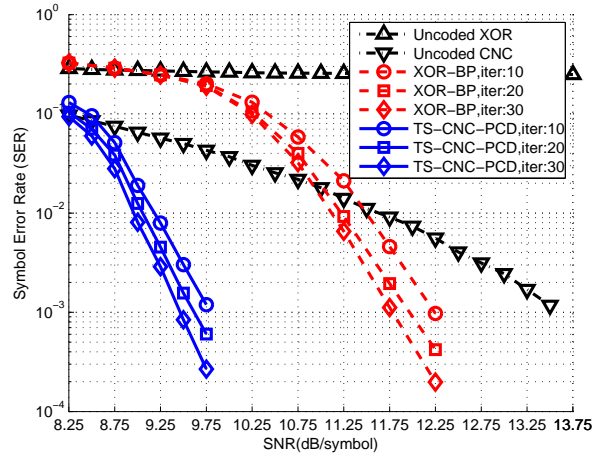}
\label{c}}} \caption{Convergence behaviors of the proposed PCD
algorithm.} \label{fig:relay_rotate}
\end{figure}

Fig.~\ref{fig:relay_rotate} shows the error performance of the
proposed PCD algorithm over three deterministic channels whose channel coefficients are fixed. The SER plotted in simulation is defined at the relay node over the MA phase only.
The number of maximum iterations is set as 10, 20 and
30. In general, larger maximum iteration leads to better performance
for the proposed TS-CNC-PCD method no matter which channel gain pair
is considered, as shown in Fig.~\ref{fig:relay_rotate}.

In Fig.~\ref{fig:relay_rotate}(a) where $H_{AC} = H_{BC}=1$, it is observed that uncoded XOR and CNC have the same performance. It is because the CNC mapping is degraded to XOR mapping when the two channel gains $H_{AC}$ and $H_{BC}$ are identical. Interestingly, the proposed TS-CNC-PCD is a little bit better than the XOR-BP. Since the traditional XOR can work well in this case, this observation confirms that there is no inherent loss coupling this non-linear operation with a LDPC code (i.e. TS-CNC-PCD) compared to the referred linear operation XOR-BP. Due to that the non-linear operation usually result in shattered pairwise check constraints. In other words, it decreases the Hamming distance of generated codeword space. Fortunately, the proposed TS-CNC-PCD not only compensates the aforesaid performance loss but also obtains extra improvement, through enlarging the cardinality of the decoding symbols and increasing the symbol distance between two distinct PLNC mapped symbols. It is clear that the coding gains of all considered coded systems are more than 3.5dB compared to the
uncoded scenarios at SER = $10^{-3}$.

From Fig.~\ref{fig:relay_rotate}(b), we can see that the conventional XOR mapping does not work anymore when $H_{AC}=1$ and $H_{BC}=j$. The reason is that in this case the XOR mapping badly decreases the symbol distance between two distinct PLNC mapped symbols although it has no loss in Hamming distance of generated codeword space. Fortunately, the coding gains
of TS-CNC-PCD is also more than 3.5dB compared to the uncoded CNC at
SER = $10^{-3}$.

Lastly, as shown in Fig.~\ref{fig:relay_rotate}(c) where $H_{AC}=1$ and $H_{BC}=(1+j)/2$,
the XOR-BP coding method outperforms the uncoded CNC more than
1.25dB at SER = $10^{-3}$ while the uncoded XOR mapping still does
not work well. At the same time, the proposed TS-CNC-PCD obtains
more than 2dB coding gain compared to the XOR-BP at SER = $10^{-3}$.

In general, our proposed TS-CNC-PCD not only obtains the best
performance but also achieves the constant coding gain in considered
TWR deterministic channels.

\subsection{Block fading channels}
\begin{figure}
\centering
\includegraphics[width=3.4in]{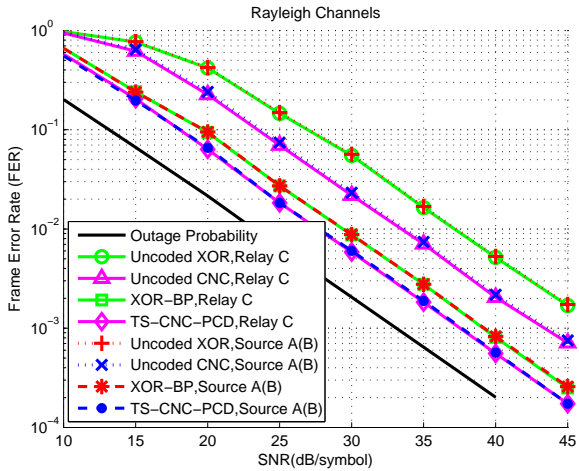}
\caption{Performance comparisons in the TWR block fading channels.}
\label{fig:TWR_fading}
\end{figure}

Suppose that the channel gains on all links follow Rayleigh
distribution and are independent. We assume $E[|H_{AC}|^2] =
E[|H_{BC}|^2] = 1$, where notation $E[\cdot]$ denotes expectation
function. The black solid lines, denoted as "Outage Probability",
are actually the lower bound of the outage probability of the TWR
block fading channels according to (\ref{eq:outage_lower_bound}).
The maximum number of iterations is fixed at 30 for the coded cases.

Fig.~\ref{fig:TWR_fading} shows the frame error rate (FER) performance of the Rayleigh
channels. For the uncoded cases, the CNC outperforms the XOR about 4
dB at FER = $2\times 10^{-3}$. At the same FER, the coding gain of
the coded XOR is about 8.5 dB. Moreover, the coding gain of the
coded CNC is about 6 dB at FER = $2\times 10^{-3}$. The reason that the coding
gain of TS-CNC-PCD is less than that of XOR-BP is that the
uncoded denoising mapping used at the relay node can eliminate part
of the noise, as confirmed in~\cite{JSAC09:Koike-Akino}. Nevertheless the TS-CNC-PCD still outperforms the XOR-BP about 2 dB at FER =$2\times 10^{-3}$. At the same time, the gap between the TS-CNC-PCD and the lower bound of the outage probability is 4.5 dB. By any possibility, we can reduce the gaps by increasing the code length.

Lastly, it is important to see that both the relay and the sources
obtain the same FER performance. These phenomena confirm that the
errors at the relay have severe impact on the performance of whole
system. This further confirms the needs for advantage decoding at the relay, such as the proposed PCD algorithm.

\section{Conclusion}

In this paper, maintaining the traditional channel coding structure
at the two sources, we propose a general relay decoding framework, called \emph{pairwise check decoding}, or \emph{PCD}, for any given PLNC mapping.
The check-relation-tab for the superimposed LDPC-coded packet pair at relay is formed according to
the selected PLNC mapping. Our proposed PCD algorithm
is universal for any adaptive PLNC mapping. In order to optimize the MED, we also present a
partial decoding method at the relay, TS-CNC-PCD, for LDPC coded TWR
block fading channels. Moreover, a check-relation-tab optimization
MCO is introduced to improve performance. Simulation results confirm
that the proposed TS-CNC-PCD has significant coding gains compared to the conventional XOR with BP and the uncoded system for certain TWR deterministic channels. For TWR fading channels, the TS-CNC-PCD also considerably outperforms
the conventional XOR with BP.


\bibliographystyle{IEEEtran}
\bibliography{IEEEabrv,reference}

\begin{IEEEbiography}[{\includegraphics[width=1in,height=1.25in,clip,keepaspectratio]{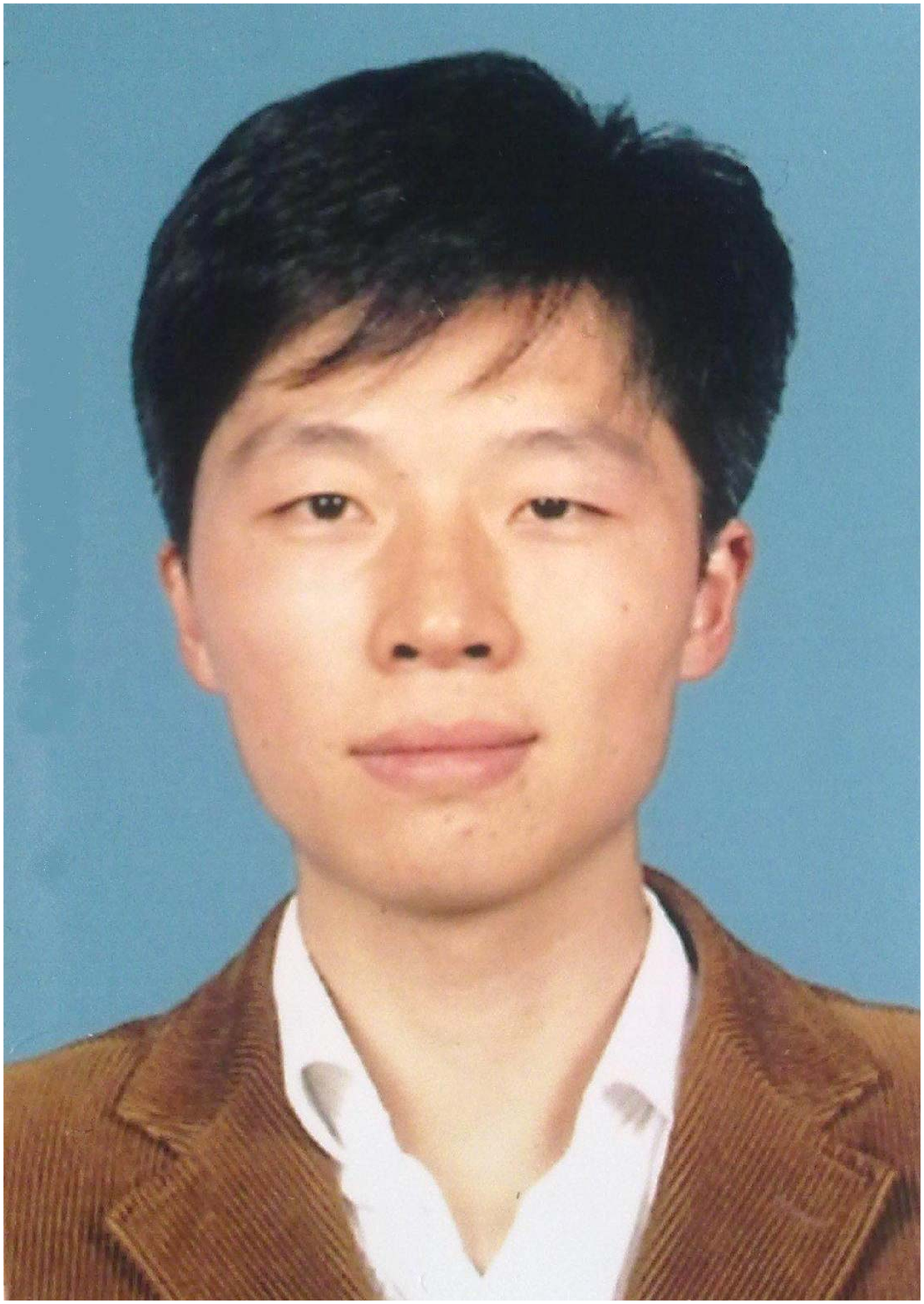}}]{Jianquan Liu}
received the B.S. degree in electronics engineering, and the M.S. degree in communication and information systems from PLA University of Science \& Technology (PLAUST), Nanjing, China, in 2000 and 2007, respectively. He is working towards his Ph.D. degree in information and communication engineering at Shanghai Jiao Tong University (SJTU), Shanghai, China. His research interests include Channel Coding, Two-Way Relaying, Physical Layer Network Coding.
\end{IEEEbiography}

\begin{IEEEbiography}[{\includegraphics[width=1in,height=1.25in,clip,keepaspectratio]{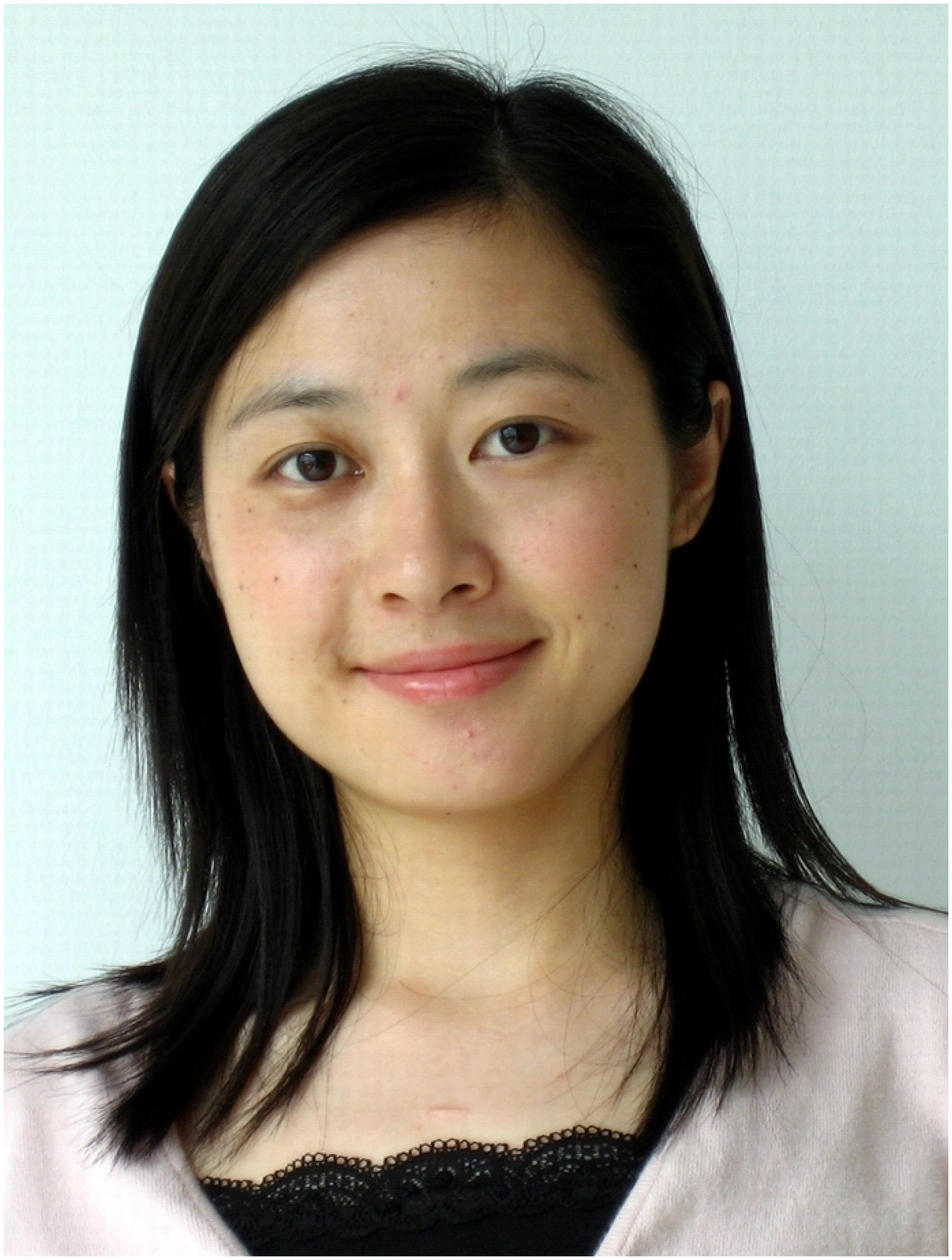}}]{Meixia Tao}
(S'00-M'04-SM'10) received the B.S. degree in electronic engineering from Fudan University, Shanghai, China, in 1999, and the Ph.D. degree in electrical and electronic engineering from Hong Kong University of Science \& Technology in 2003. She is currently an Associate Professor at the Department of Electronic Engineering, Shanghai Jiao Tong University, China. From Aug. 2003 to Aug. 2004, she was a Member of Professional Staff at Hong Kong Applied Science \& Technology Research Institute Co. Ltd. From Aug 2004 to Dec. 2007, she was with the Department of Electrical and Computer Engineering at National University of Singapore as an Assistant Professor. Her current research interests include cooperative transmission, physical layer network coding, resource allocation of OFDM networks, and MIMO techniques.

Dr. Tao is an Editor for the IEEE WIRELESS COMMUNICATIONS LETTER, and an Associate Editor for the IEEE COMMUNICATIONS LETTERS. She was on the Editorial Boards of the IEEE TRANSACTIONS ON WIRELESS COMMUNICATIONS from 2007 to 2011 and the JOURNAL OF COMMUNICATIONS AND NETWORKS from 2009 to 2011. She served as Track/Symposium Co-Chair for APCC09, ChinaCom09, IEEE ICCCN07, and IEEE ICCCAS07. She has also served as Technical Program Committee member for various conferences, including IEEE INFOCOM, IEEE GLOBECOM, IEEE ICC, IEEE WCNC, and IEEE VTC.

Dr. Tao is the recipient of the IEEE ComSoC Asia-Pacific Outstanding Young Researcher Award in 2009.

\end{IEEEbiography}

\begin{IEEEbiography}[{\includegraphics[width=1in,height=1.25in,clip,keepaspectratio]{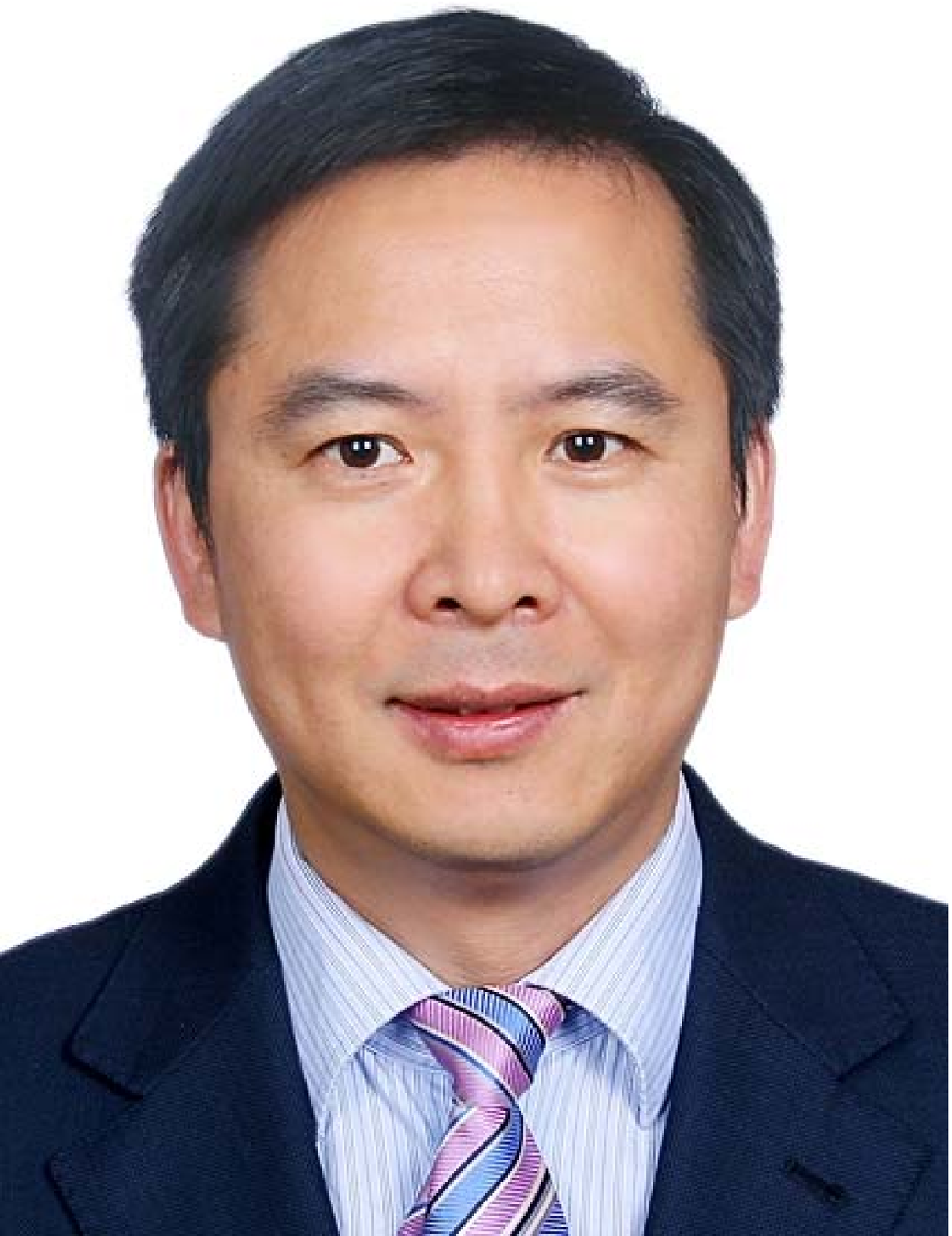}}]{Youyun Xu}
received the Ph.D. Degree in information and communication engineering from Shanghai Jiao Tong University, Shanghai, China, in 1999. He is currently a professor in Nanjing Institute of Communication Engineering, PLA University of Science \& Technology (PLAUST), China. He is also a part-time professor with the Institute of Wireless Communication Technologies of Shanghai Jiao Tong University (SJTU), China. He has more than 20-year professional experience of teaching and researching in communication theory and engineering. Now, his research interests are focusing on New Generation Wireless Mobile Communication System (LTE£¬IMT-Advanced and Related), Advanced Channel Coding and Modulation Techniques, Multiuser Information Theory and Radio Resource Management, Wireless Sensor Networks, Cognitive Radio Networks, etc. He is a senior member of the IEEE and a senior member of Chinese Institute of Electronics.
\end{IEEEbiography}

\end{document}